\documentclass[conference]{IEEEtran}
\IEEEoverridecommandlockouts
\usepackage[table]{xcolor}
\usepackage{cite}
\usepackage{amsmath,amssymb,amsfonts}
\usepackage{amsbsy}
\usepackage{graphicx}
\usepackage{textcomp}
\usepackage{xcolor}
\def\BibTeX{{\rm B\kern-.05em{\sc i\kern-.025em b}\kern-.08em
    T\kern-.1667em\lower.7ex\hbox{E}\kern-.125emX}}
    
\usepackage[utf8]{inputenc}
\usepackage{url}
\usepackage{amsmath}
\usepackage{mathtools,amssymb,lipsum,nccmath,bm,amsmath}

\usepackage{cuted}
\usepackage{array}
\usepackage{ragged2e}
\usepackage{optidef}
\usepackage{algorithm,algcompatible}
\usepackage{siunitx}

\algnewcommand\INPUT{\item[\mathbf{Inputs:}]}
\algnewcommand\OUTPUT{\item[\mathbf{Output:}]}
    
\usepackage{dblfloatfix}
\usepackage{float}
  \floatstyle{ruled}
  \newfloat{algorithm}{htbp}{loa}
  \floatname{algorithm}{Algorithm}
\usepackage{url}

\usepackage{balance}%
\usepackage{amsthm}

\newtheorem*{remark}{Remark}

\begin{document}

\title{Online Dictionary Learning Based Fault and \\ Cyber Attack Detection for Power Systems}


\author{\IEEEauthorblockN{Gabriel Intriago and Yu Zhang}
\IEEEauthorblockA{Department of Electrical and Computer Engineering \\
University of California, Santa Cruz\\
Emails:  \{gintriag, zhangy\}@ucsc.edu}
\thanks{This work was supported in part by a Seed Fund Award from CITRIS and the Banatao Institute at the University of California, and the Hellman Fellowship.}}

\maketitle

\begin{abstract}
The emerging wide area monitoring systems (WAMS) have brought significant improvements in electric grids' situational awareness. However, the newly introduced system can potentially increase the risk of cyber-attacks, which may be disguised as normal physical disturbances. This paper deals with the event and intrusion detection problem by leveraging a stream data mining classifier (Hoeffding adaptive tree) with semi-supervised learning techniques to distinguish cyber-attacks from regular system perturbations accurately. First, our proposed approach builds a dictionary by learning higher-level features from unlabeled data. Then, the labeled data are represented as sparse linear combinations of learned dictionary atoms. We capitalize on those sparse codes to train the online classifier along with efficient change detectors. We conduct numerical experiments with industrial control systems cyber-attack datasets. We consider five different scenarios: short-circuit faults, line maintenance, remote tripping command injection, relay setting change, as well as false data injection. The data are generated based on a modified IEEE 9-bus system. Simulation results show that our proposed approach outperforms the state-of-the-art method.

\end{abstract}


\section{Introduction}

The ongoing improvements in wide-area monitoring systems have brought better visibility of the power system and have exposed the system to malicious cyber-attacks \cite{Ntalampiras2015}, \cite{Lin2012}. In this context, event and intrusion detection systems (EIDS) are indispensable to classify the nature of a power system disturbance: is it a regular operation, fault condition, or a cyber-attack? The main challenge of this classification task is to extract relevant information from system measurements. Over the past decade, various data-driven techniques have been explored to tackle this problem. 

A geometrical analysis of unsynchronized and synchronized attacks is introduced to detect the presence of attacks and identify compromised micro-PMUs \cite{Kamal2020}. Based on text-mining techniques, a data-driven approach was developed for false data attacks classification \cite{Ma2020}. In recent years, classical machine learning algorithms, such as naive Bayes, support vector machines (SVM), and random forests (RF), have been applied to detect cyber-attacks and disturbances in power systems \cite{BorgesHink2014}, \cite{AlKarim2012}, \cite{Demertzis2018}, \cite{Wang2019}. It is also possible to build common paths of critical states by exploiting the relationships among voltage, current, and impedance to discover relevant patterns \cite{Pan20152}, \cite{Pan2015}. Those classical methods, which often have difficulties dealing with large-scale and time-varying data, are unsuitable for real-time changing environments. Hence, stream data mining algorithms have recently drawn much attention. These include nonnested generalized exemplars \cite{Adhikari2018},  Hoeffding adaptive tree (HAT) \cite{Dahal2015}, and HAT with change detectors \cite{Adhikari20182}. Leveraging phasor measurement units (PMUs), those algorithms are proven to outperform classical methods in real systems.  In \cite{Mrabet2019}, the authors proposed a transfer learning HAT model with one change detector, the adaptive sliding window (ADWIN). Their approach transferred knowledge from four datasets, where each dataset corresponds to a specific frequency oscillation.

It is challenging and costly to label a massive amount of PMU data on the fly in practice. Compared with data collection that depends only on data storage capacity, data labeling often requires rich domain knowledge of experts who can actively identify instances' labels. Therefore, we have abundant unlabeled data and scarce labeled data that share the same generative distribution. Due to this fact, semi-supervised learning (SSL) is an appropriate tool that combines a small amount of labeled data with a large amount of unlabeled data during training \cite{Nigam2000}.

This paper proposes a novel approach for power system EIDS to improve the classification performance by transforming the data through higher-level representations extracted from an unlabeled dataset. In addition, we provide performance analysis for different sizes of the labeled dataset. To the best of our knowledge, this is the first effort to incorporate SSL with a stream data mining classifier for the EIDS. The rest of the paper is organized as follows. Section II presents the details of the proposed approach. Section III shows the simulation results. Finally, section IV gives the conclusion.




\section{Semi-Supervised Hoeffding Adaptive Tree}
We learn a dictionary by extracting higher-level features (such as oscillations, sudden changes, gradual changes, stable periods) from the unlabeled dataset to represent later the labeled data, which are then used to train a classifier incrementally.

\subsection{Online Dictionary Learning}
Given a set of unlabeled instances $\mathcal{U} = \left\{\mathbf{x}_{\text{u}}^{(1)},  \ldots, \mathbf{x}_{\text{u}}^{(p)} \right\}$, where $\mathbf{x}_{\text{u}}^{(i)} \in \mathbb{R}^n$ is the $i$-th input feature vector, we formulate the following  optimization problem to learn a new feature space representing these data points:
\begin{mini!}
  {\mathbf{D}, \{\bm{\alpha}_{\text{u}}^{(i)}\}}{\frac{1}{2}\sum_{i=1}^{p}\left\|\mathbf{x}_{\text{u}}^{(i)} - \mathbf{D}\bm{\alpha}_{\text{u}}^{(i)}\right\|_2^2}{}{}
  \addConstraint{\left\|\bm{\alpha}_{\text{u}}^{(i)}\right\|_0 \leq k,\,  i = 1,2,\ldots,p}\label{constr:sparse}
  \addConstraint{\left\|\bm{d}_{j}\right\|_2 \leq 1,\,   j = 1,2,\ldots,m}
\label{constr:unitnorm}
 \end{mini!}
The optimization variables are the dictionary $\mathbf{D}_{t}$ = $\big[\mathbf{d}_1, \ldots, \mathbf{d}_m\big] \in \mathbb{R}^{n\times m}$ and the sparse codes $\bm{\alpha}_{\text{u}}^{(i)}$ $\in \mathbb{R}^{m}, i =1, 2, \ldots, p$.
Typically, we have $m\gg n$ so that the dictionary is rich enough. Hence, by the least square objective, each input $\mathbf{x}_{\text{u}}^{(i)}$ is approximately represented as a linear combination of very few basis vectors in $\mathbf{D}$ with the corresponding coefficients given by $\bm{\alpha}_{\text{u}}^{(i)}$.
The zero norm $\|\mathbf{a}\|_0$ denotes the number of non-zero coordinates of $\mathbf{a}$. Hence, the first constraint forces the vector $\bm{\alpha}_{\text{u}}^{(i)}$ to have at most $k$ nonzero elements. 
The energy of each atom (basis) in the dictionary $\mathbf{D}$ is bounded by one, as given by the second constraint. This constraint prevents the entries of $\mathbf{D}$ from being arbitrarily large while the entries of $\bm{\alpha}_{\text{u}}^{(i)}$ being very small. 

We leverage the alternating minimization method for the resulting nonconvex problem (1), i.e., minimizing one variable at each step while keeping all other variables fixed \cite{Mairal2009}. In the first step, we obtain the sparse codes $\bm{\alpha}_{\text{u}}^{(i)}, i=1,2\ldots,p$. The second step updates the dictionary $\mathbf{D}$. 

\begin{itemize}
\item \textit{Sparse coding -- optimization over} $\bm{\alpha}_{\text{u}}^{(i)}$: 
Start with a fixed random dictionary $\mathbf{D}$, and solve (1) with the orthogonal matching pursuit (OMP) algorithm to obtain the $\bm{\alpha}_{\text{u}}^{(i)}$ that corresponds to the unlabeled point $\mathbf{x}_{\text{u}}^{(i)}$ for $i=1,2,\ldots,p$. 

\item \textit{Dictionary update -- optimization over} $\mathbf{D}$:
Keep $\{\bm{\alpha}_{\text{u}}^{(i)}\}_{i=1}^p$ fixed, find the dictionary $\mathbf{D}_{t}$ by sequentially updating each atom via the block-coordinate descent (BCD) algorithm:
\begin{align}
\mathbf{u}_j &= \mathbf{A}_{jj}^{-1}(\mathbf{b}_j - \mathbf{D}_{t-1}\mathbf{a}_j) + \mathbf{d}_j,\, &j=1,2,\ldots, m \\
\mathbf{d}_j &= \frac{\mathbf{u}_j}{\max(\|\mathbf{u}_j\|_2,1)},\, &j=1,2,\ldots, m, 
\end{align}
where $\mathbf{D}_{t-1}$ is the dictionary at the previous iteration. 
The matrices $\mathbf{A} = \big[\mathbf{a}_1,\ldots, \mathbf{a}_m\big] = \bm{\alpha}_{\text{u}}^{(i)}\bm{\alpha}_{\text{u}}^{(i)^{\top}} \in \mathbb{R}^{m\times m}$ and $\mathbf{B} = \big[\mathbf{b}_1, \ldots, \mathbf{b}_m\big] = \mathbf{x}_{\text{u}}^{(i)}\bm{\alpha}_{\text{u}}^{(i)^{\top}}  \in \mathbb{R}^{n\times m}$  carry the information of the updated $\bm{\alpha}_{\text{u}}^{(i)}$'s. The update repeats until $\mathbf{D}_{t}$ converges.
\end{itemize}

\subsection{New Feature Representation}\label{nfr}
Consider a set of labeled instances $\mathcal{L} = \big\{(\mathbf{x}_{\ell}^{(1)},y^{(1)}),   \ldots, (\mathbf{x}_{\ell}^{(q)},y^{(q)}) \big\}$, where $\mathbf{x}_{\ell}^{(i)} \in \mathbb{R}^n$ is the $\ell$-th input feature vector with label $y^{(i)} \in \{1, \ldots , C \}$. Upon learning the dictionary $\mathbf{D}^{*}$ as elaborated above, the labeled data can be represented by using the basis vectors of $\mathbf{D}$. This is carried out by solving the following  problem via the OMP for each labeled data point:
\begin{mini!}
  {\bm{\alpha}_{\ell}^{(i)}}{\frac{1}{2}\left\|\mathbf{x}_{\ell}^{(i)} - \mathbf{D}^{*}\bm{\alpha}_{\ell}^{(i)}\right\|_2^2}{}{}
  \addConstraint{\left\|\bm{\alpha}_{\ell}^{(i)}\right\|_0 \leq k.}
  \label{eq2}
 \end{mini!}
In other words, a labeled data point is now approximately represented as a linear combination of the learned atoms as:
\begin{align}
\mathbf{x}_{\ell}^{(i)} = \mathbf{D}\bm{\alpha}_{\ell}^{(i)} + \bm{\eta},
\end{align}
where $\bm{\eta}$ is the reconstruction error. 
We preserve each original label of $y^{(i)}$ by attaching it to the new representation; i.e., the $k$-sparse code $\bm{\alpha}_{\ell}^{(i)}$ in a higher dimensional space.
Finally, we train the HAD classifier with these new representations by using the software package MOA \cite{Bifet2018}.


\begin{algorithm}[t]
    \caption{Semi-supervised HAD (SSHAD)}
  \begin{algorithmic}[1]
  \RaggedRight
    \REQUIRE \phantom{This text will be invisible}\linebreak
    1) Unlabeled data $\mathcal{U}$ = $\{\mathbf{x}_{\text{u}}^{(1)}, \ldots, \mathbf{x}_{\text{u}}^{(p)} \}$\linebreak
    2) Labeled data $\mathcal{L}$ = $\big\{(\mathbf{x}_{\ell}^{(1)},y^{(1)}),\ldots (\mathbf{x}_{\ell}^{(q)},y^{(q)}) \big\}$.\linebreak
    3) Randomly initialize $\mathbf{D}$ from unlabeled data vectors.\linebreak
    4) Maximum iteration: $\text{max\_iter} = 200$.
    \STATE Normalize the labeled and unlabeled data.
    \FOR{$t= 1$ to \text{max\_iter}}
      \STATE Compute the sparse code $\bm{\alpha}_{\textrm{u}}$ with $\mathbf{D}_{t-1}$ by solving (1).
      \STATE Update $\mathbf{D}_{t}$ keeping the matrix $\bm{\alpha}_u$ fixed.
    \ENDFOR
      \STATE Solve \eqref{eq2} to obtain the matrix $\bm{\alpha}_\ell$.
      \STATE Attach to $\bm{\alpha}_\ell$ the labels from $\mathbf{x}_{\ell}$.
    \STATE Train HAD with the new labeled dataset $\hat{\mathcal{L}} = \left\{ (\bm{\alpha}_\ell^{(1)},y^{(1)}), \ldots, (\bm{\alpha}_\ell^{(q)},y^{(q)}) \right\}$ using MOA.
    \STATE $\textbf{return}$ The trained HAD classifier.
  \end{algorithmic}
  \label{alg1}
\end{algorithm}

\begin{remark}[Matching pursuit vis-a-vis LASSO]
The sparse dictionary learning problem generally has two different formulations: matching pursuit and LASSO. The former is shown by problem (1) while the latter is relaxing $\ell_0$ norm to $\ell_1$ norm and being lifting to the objective as a soft constraint.
The matching pursuit formulation explicitly guarantees $k$-sparsity, which is more user-friendly to find the ``best'' value of $k$ by trial-and-error simulations. According to our numerical experiments that will be discussed in the next section, we find that the solution to the matching pursuit is more stable numerically.  
\end{remark}

Algorithm~\ref{alg1} features two essential differences from the algorithm in \cite{Liu2018}. In \cite{Liu2018}, the authors build the dictionary using self-taught learning (unlabeled and labeled datasets have different generative distributions \cite{Raina2007})  to later train and test an SVM classifier with the new representation of the labeled dataset. In contrast, our model builds the dictionary using SSL and next incrementally trains a HAD classifier with all the transformed labeled dataset instances. In a nutshell, our algorithm capitalizes on semi-supervised knowledge to enhance the HAD classifier's overall performance. We name the proposed algorithm as SSHAD, where ``SS''  stands for semi-supervised, to differentiate it from the original version of HAD presented in \cite{Adhikari20182}.

\subsection{HAD Classifier}

HAD is composed of three main ingredients: a window to remember recent examples, a distribution-change detector, and an estimator for some statistics of the input data. Once a change is detected, an alternate tree will be created and grow with the instances appearing right after the change. The existing alternate tree will replace the current tree if it is more accurate. The HAT \cite{Bifet2009} is the parent tree of HAD, where the former has only one change detector, ADWIN, whereas HAD has two change detectors ADWIN and DDM.

ADWIN serves as an estimator and change detector that keeps a variable-length window $\mathcal{W}$ of recent data such that the window has the maximal length statistically consistent with the null hypothesis of the average value inside the window has not changed. When two ``big enough" sub-windows of $\mathcal{W}$ have ``distinct enough" averages, it can be said with high probability that a change in the data distribution has occurred and the older items in $\mathcal{W}$ should be dropped. The ``big and distinct enough'' can be quantitatively defined by the Hoeffding bound \cite{Bifet2007}.

DDM is a change detector that relies on the concept of `context' defined as a set of contiguous examples whose data distribution is stationary. DDM incrementally controls the error rate of the model. Statistical theory guarantees that the error decreases if the data distribution remains stationary, and error increases when the distribution changes. A new context is declared if the error reaches a warning level at instance $k_w$ and a drift level at instance $k_d$. Given that, this indicates a distribution change, and a new model is learned by using the examples between $k_w$ and $k_d$. A detailed explanation of DDM can be found in \cite{Gama2004}.

\section{Experiments and Results}
\subsection{Datasets}
Power system attack datasets \cite{datasets} are used to test the performance of our proposed approach. There are three datasets: 2-class, 3-class, and 37-class datasets, where each of them includes 128 features split into two categories: physical (voltages, currents, and impedances) and cyber-physical (control logs, network alerts, and relay logs) features. Five scenarios are considered: short-circuit faults, line maintenance, remote tripping command injection (attack), relay setting change (attack), as well as data injection (attack). Fig.~\ref{fig7} shows the testbed architecture used in generating the datasets.


 \begin{figure}
     \centering
     \includegraphics[width=1\linewidth]{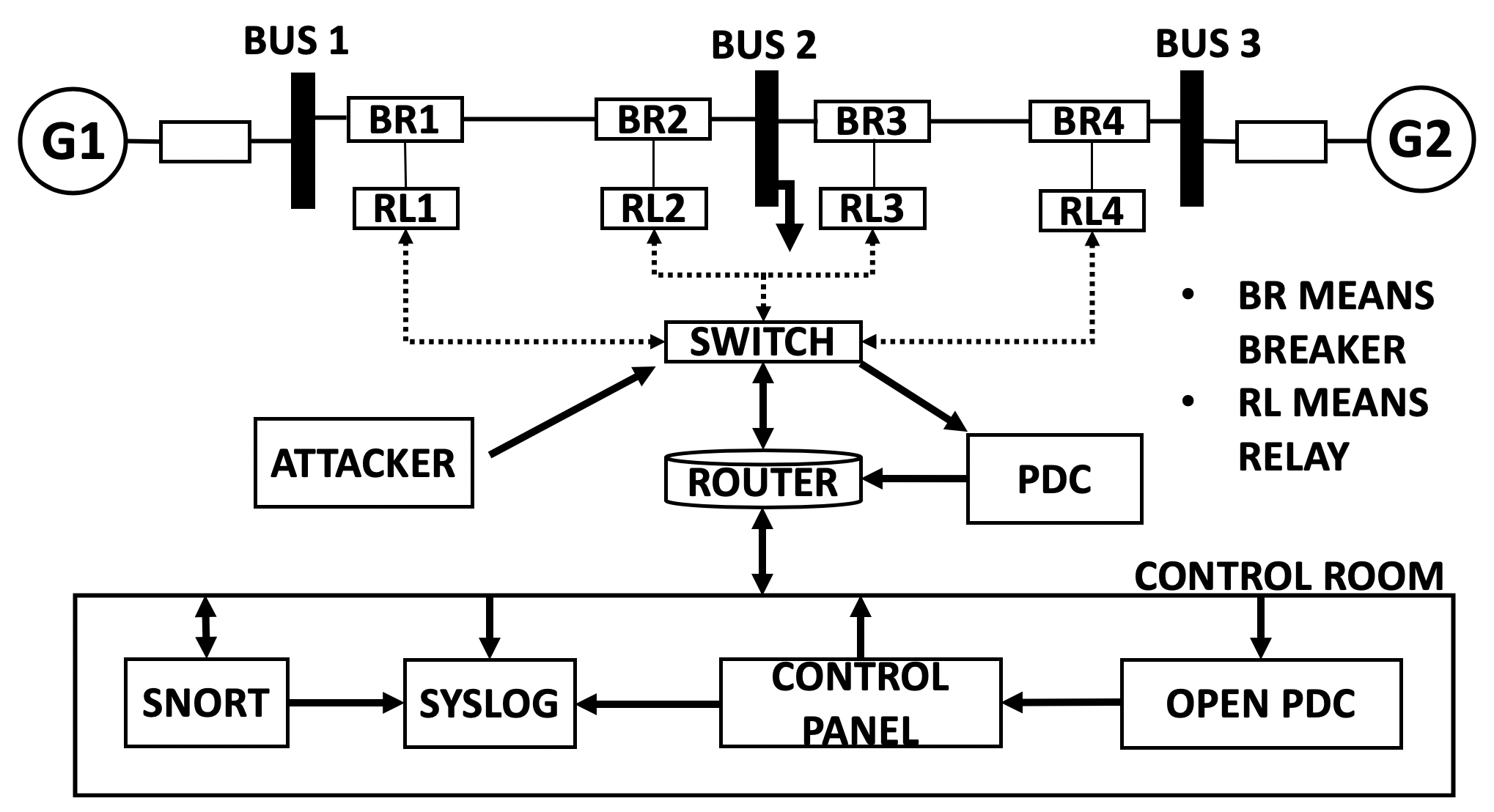}
     \caption{Test-bed architecture for generating the datasets}
     \label{fig7}
 \end{figure}

\subsection{Implementation and Parameters}

We run all the experiments using MATLAB, WEKA, and the massive online analysis (MOA) software \cite{Bifet2014}. The relevant parameters were obtained by using cross-validation. The value of $\text{max\_iter}=200$ yielded best results. The parameter $k$ was set to 10 for both OMP procedures, i.e, each of $\bm{\alpha}_{\textit{u}}^{(i)}$'s and $\bm{\alpha}_{\ell}^{(i)}$'s has at most ten nonzero values. 
We tested different sizes for the dictionary and found that 130 atoms performed the best. For both OMP optimization problems, the tolerance of the squared $\ell_2$-norm residual was set to 0.01. Finally, the parameters for the HAD were set to the default values given by MOA. 

\subsection{Performance Metrics}
In this work, we used the prequential evaluation technique, where each instance is used to test and then train the model. Because of this online setup, the accuracy is incrementally updated. We chose the classification accuracy, the Kappa statistic, evaluation time, and model cost to evaluate our approach's performance; see also~\cite{Adhikari20182}. The Kappa statistic is a measure for rating classification accuracy for imbalance scenarios in offline and online classification. The Kappa statistic is defined as:
\begin{align} 
		\kappa &= \frac{\rho_o - \rho_e}{1 - \rho_e},
\end{align}
where $\rho_o$ is the accuracy of the classifier under analysis, and $\rho_e$ is the accuracy of a random classifier. If the classifier predicts all the time correctly, $\kappa = 1$. If the classifier performs like a random classifier, $\kappa = 0$. The evaluation time consists of both training and testing time because there is no clear separation between them \cite{Adhikari20182}. The model cost is measured in RAM per hour (hereafter referred to as Ram-Hours)~\cite{Dahal2015}. 



\subsection{Simulation Results}
We conduct classification experiments using the 2-class, 3-class, and 37-class datasets. The performance results were obtained with five different sizes, determined by the labeled dataset's sampling ratio. All values given in figures and tables are 10-fold average. The performance of our model improves with the increased size of the unlabeled dataset. It can be seen that the performance gets saturated with 50,000 unlabeled data points.

\begin{figure}[H]
     \centering
     \includegraphics[width=0.95\linewidth]{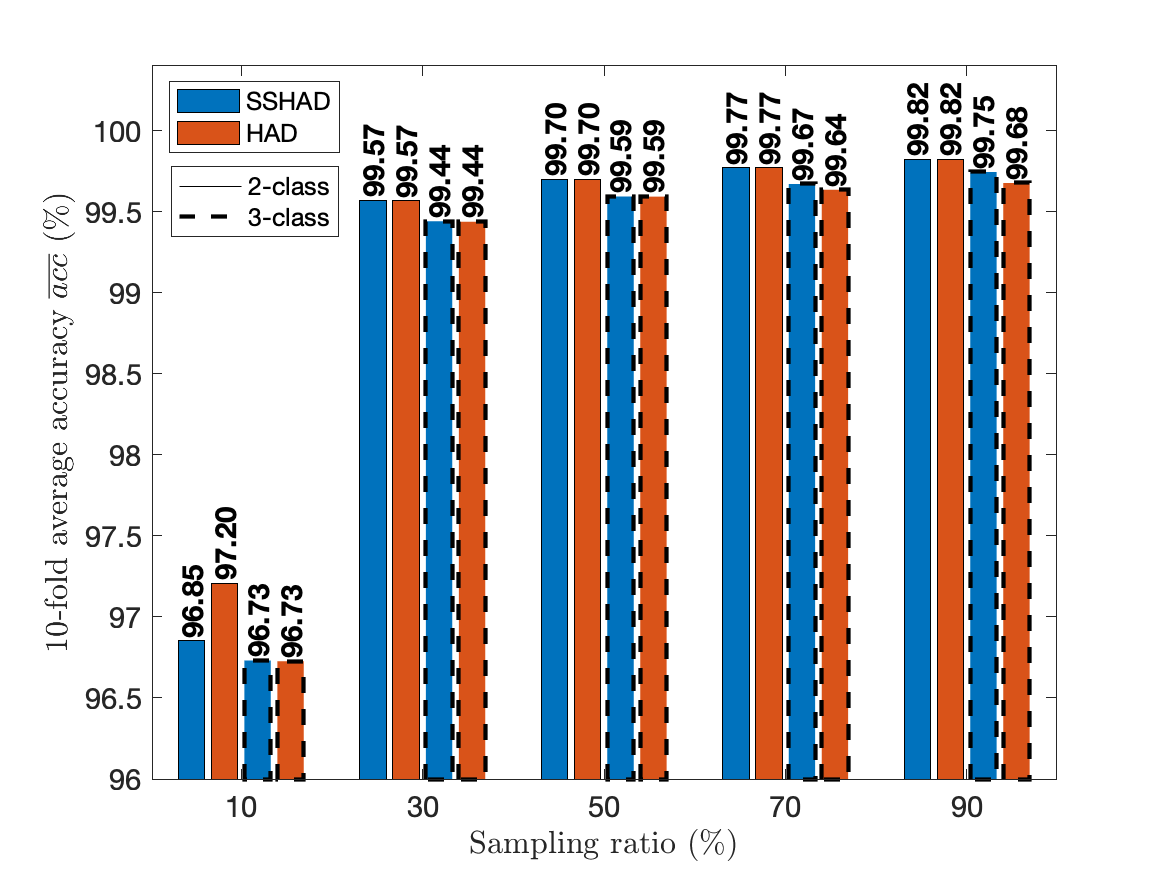}
     \caption{10-fold average accuracy $(\overline{acc})$ comparison between SSHAD with parameter $k$ = 10 and HAD using the 2-class and 3-class datasets.}
     \label{fig5}
 \end{figure}

\begin{figure}[H]
     \centering
     \includegraphics[width=0.95\linewidth]{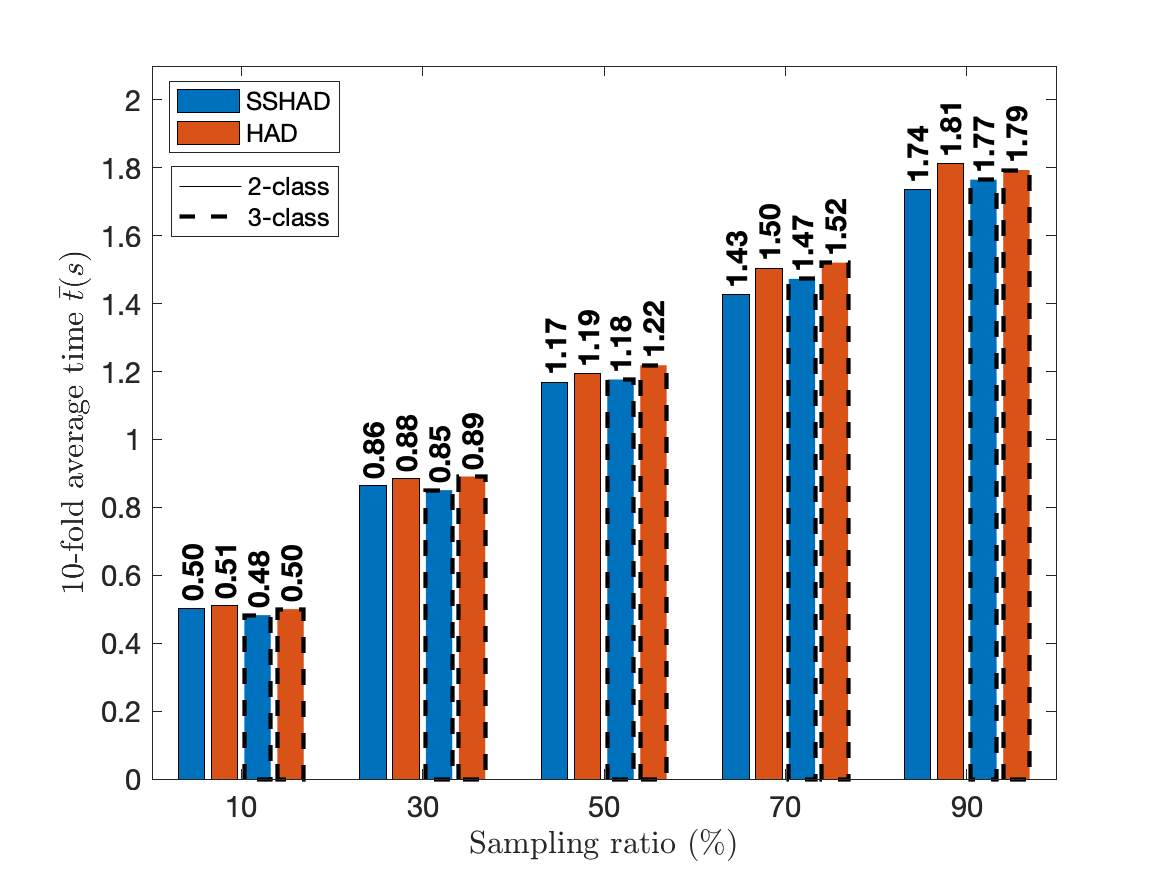}
     \caption{10-fold average accuracy $(\overline{acc})$ comparison between SSHAD with parameter $k$ = 10 and HAD using the 2-class and 3-class datasets.}
     \label{fig3}
 \end{figure}

\begin{figure}[H]
     \centering
     \includegraphics[width=0.95\linewidth]{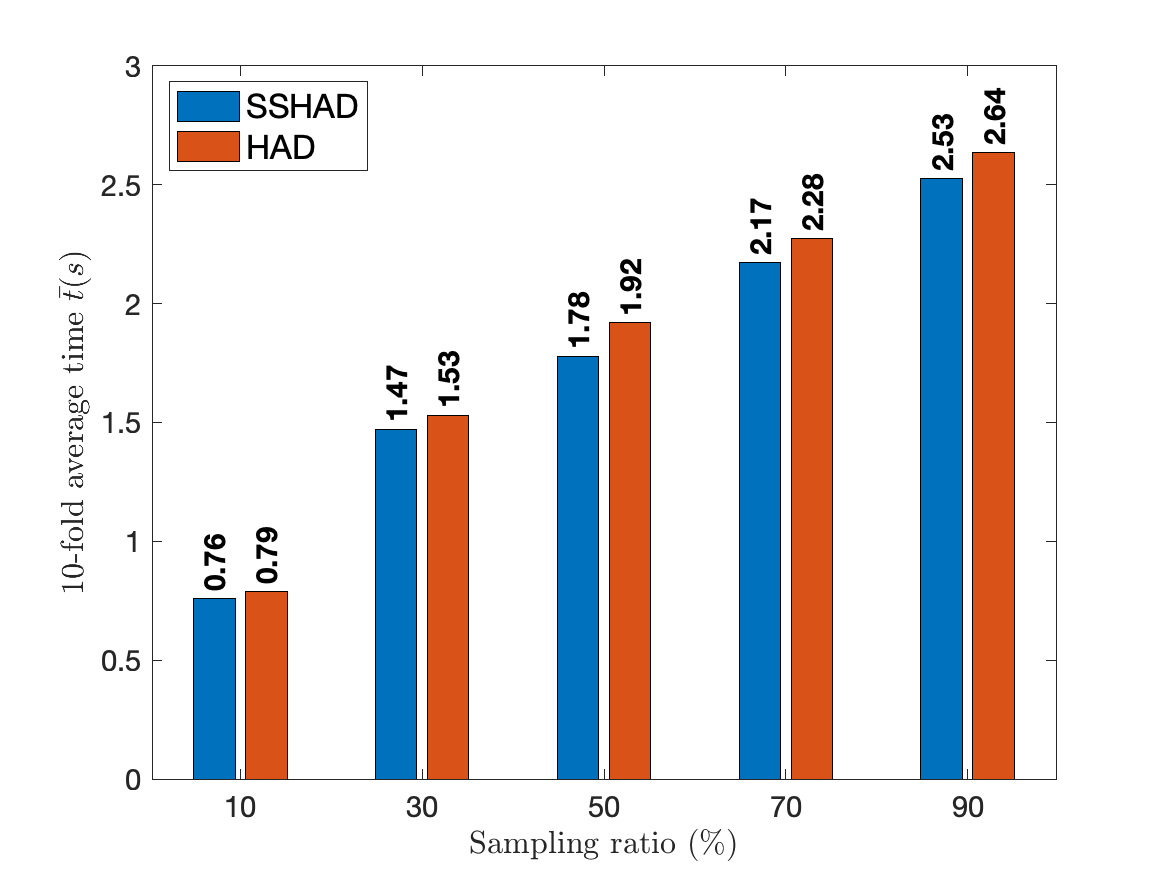}
     \caption{10-fold average time ($\bar{t}$) comparison between SSHAD with parameter $k$ = 10 and HAD using the 37-class dataset.}
     \label{fig2}
     \vspace{-2cm}
 \end{figure}

Fig. \ref{fig5}, \ref{fig3} and Tab. \ref{table1} show the classification results for the 2-class and 3-class datasets. It can be seen that the performances of SSHAD and HAD are similar. However, when it comes to the 37-class dataset, our model clearly outperforms HAD as shown in Fig. \ref{fig1}, \ref{fig2} and Tab. \ref{table2}.  These results corroborate the merits of our proposed approach, representing the data by higher-level features yields more accurate identification of events in power systems. Moreover, as shown in Fig. \ref{fig7}, SSHAD is robust to the presence of bad data.

 \begin{table}[H]
\caption{The 3-class dataset: 10-fold average Kappa $(\overline{\kappa})$ and cost $(\overline{cost})$ comparisons between SSHAD ($k$ = 10) and HAD.}
\begin{center}
\begin{tabular}{c||c|c|c|c} 
 \hline
Sampling&\multicolumn{2}{c|}{$\overline{\kappa}({\%})$}&\multicolumn{2}{c}{$\overline{cost}$~(Ram-Hour)}\\
 \cline{2-5}
Ratio& SSHAD &HAD & SSHAD &HAD\\
 \hline
 10\%&82.25&82.29& \num{1.43e-08}&\num{1.37e-08}\\
 \hline
 30\%&88.36&88.44& \num{2.18e-08}&\num{2.07e-08}\\ 
 \hline
 50\%&\textbf{69.87}& 69.57&\boldsymbol{$2.87\times 10^{-8}$}&\num{2.89e-08}\\ 
 \hline
 70\%&\textbf{59.56}&59.28&\boldsymbol{$3.67\times 10^{-8}$}&\num{3.83e-08}\\ 
 \hline
 90\%&\textbf{51.91}&51.70&\boldsymbol{$4.63\times 10^{-8}$}&\num{5.27e-08}\\ 
 \hline
 \end{tabular}
\end{center}
\label{table1}
\end{table}

     \begin{figure}[H]
     \centering
     \includegraphics[width=0.95\linewidth]{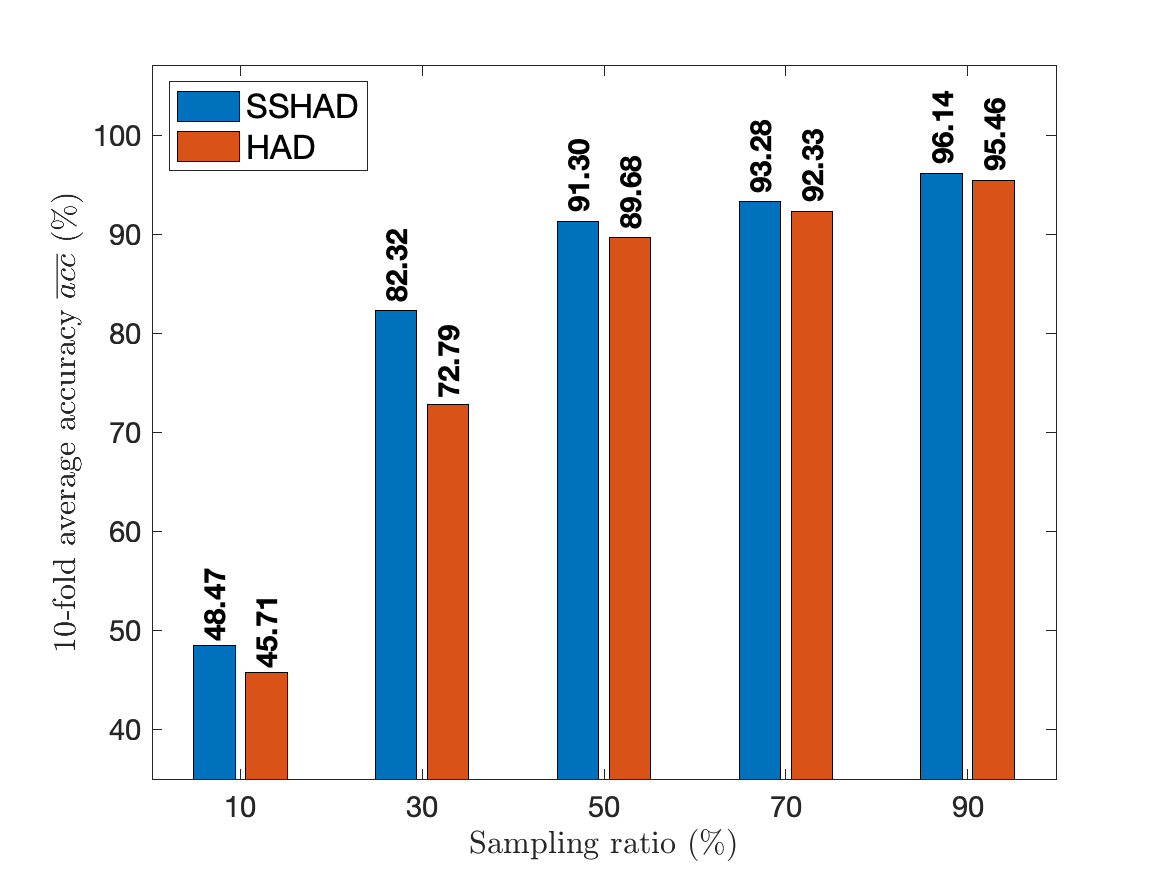}
     \caption{10-fold average time ($\bar{t}$) comparison between SSHAD with parameter $k$ = 10 and HAD using the 37-class dataset in the presence of $10\%$ of bad data.}
     \label{fig7}
     \vspace{-2cm}
 \end{figure}
 
  \begin{figure}[H]
     \centering
     \includegraphics[width=0.95\linewidth]{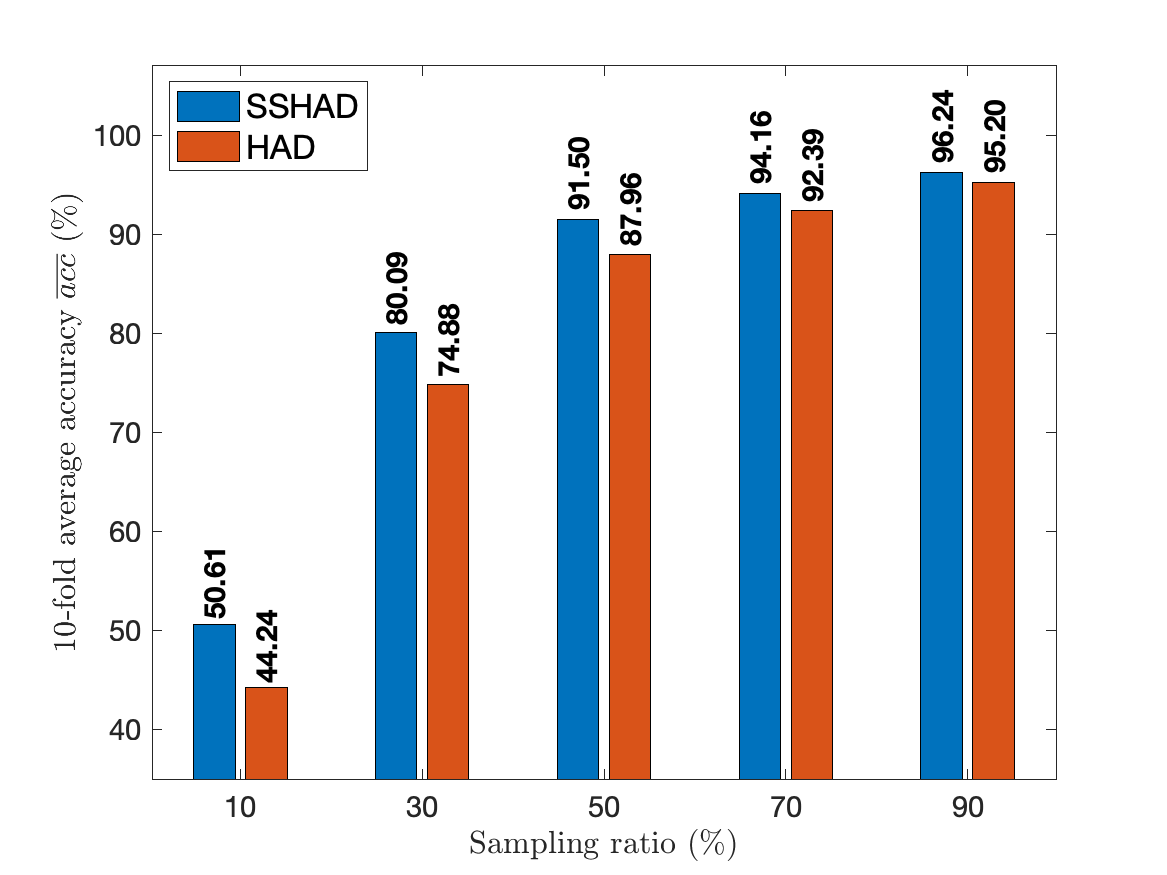}
     \caption{10-fold average accuracy $(\overline{acc})$ comparison between SSHAD with parameter $k$ = 10 and HAD using the 37-class dataset.}
     \label{fig1}
 \end{figure} 
 
 \begin{table}[H]
\caption{The 37-class dataset: 10-fold average Kappa $(\overline{\kappa})$ and cost $(\overline{cost})$ comparisons between SSHAD ($k$ = 10) and HAD.}
\begin{center}
\begin{tabular}{ c||c|c|c|c } 
 \hline
Sampling&\multicolumn{2}{c|}{$\overline{\kappa}({\%})$}&\multicolumn{2}{c}{$\overline{cost}$~(Ram-Hour)}\\
 \cline{2-5}
Ratio& SSHAD &HAD & SSHAD &HAD\\
 \hline
 10\%&\textbf{29.80}&28.39&\boldsymbol{$5.80\times 10^{-8}$}&\num{5.88E-08}\\
 \hline
 30\%&\textbf{69.62}&62.32&\boldsymbol{$1.20\times 10^{-7}$}&\num{1.64E-07}\\ 
 \hline
 50\%&\textbf{79.48}& 77.41&\boldsymbol{$9.24\times 10^{-8}$}&\num{1.09E-07}\\ 
 \hline
 70\%&\textbf{82.33}&80.30&\boldsymbol{$1.09\times 10^{-7}$}&\num{1.13E-07}\\ 
 \hline
 90\%&\textbf{85.64}&84.38&\boldsymbol{$1.24\times 10^{-7}$}&\num{1.28E-07}\\ 
 \hline
 \end{tabular}
\end{center}
\label{table2}
\end{table}

 \section{Conclusion}\label{sec:conclu}
We develop a semi-supervised online approach (SSHAD) for the power system event detection in this paper. The labeling process for a large amount of unlabeled data is often very time-consuming and costly, requiring specific domain knowledge of many experts. Considering this fact, we leverage online dictionary learning techniques to automatically build a new feature space for the labeled data examples by extracting valuable information from the unlabeled dataset. The learned sparse codes of the labeled instances become the new feature representations, based on which we train the HAD classifier.

Extensive numerical results corroborate our proposed approach's effectiveness that yields a better classification performance and compensates for the additional computational burden of learning the higher dimensional representations. Despite these results, we acknowledge that future work is needed to make our approach more robust. For instance, this work can be extended by studying how a malicious adversary can modify the data and determining the depth of its attack from the game theory perspective. Finally, a more detailed analysis of the temporal dependence of the data should be considered.



\bibliographystyle{IEEEtran}
\bibliography{references}

\end{document}